\documentclass[12pt, a4paper]{iopart}

\usepackage{iopams}  
\usepackage{graphicx}
\usepackage{braket}
\usepackage{bm}
\usepackage{dsfont}
\usepackage{booktabs}
\usepackage[cdot,squaren, thinspace,thinqspace]{SIunits}
\usepackage{bm}
\usepackage{cite}
\usepackage{color}

\begin{document}
\title[]{Characteristics of displaced single photons attained via higher order factorial moments}

\author{Kaisa Laiho$^{1,2}$, Malte Avenhaus$^{1,2}$, and Christine Silberhorn$^{1,2}$}
\address{$^{1}$Applied Physics, University of Paderborn,  Warburgerstr.~100, 33098 Paderborn, Germany.}
\address{$^{2}$Max Planck Institute for the Science of Light, Guenther-Scharowsky-Str.~1/ Bldg.~24, 91058 Erlangen, Germany.}

\ead{kaisa.laiho@uni-paderborn.de}

\begin{abstract}
We investigate up to the fourth order normalized factorial moments of free-propagating and pulsed single photons displaced in phase space in a phase-averaged manner.
Due to their loss independence, these moments offer expedient methods for quantum optical state characterization.
We examine quantum features of the prepared displaced states, retrieve information on their photon-number content  and study the reliability of the state reconstruction method used.
\end{abstract}

\vspace{2pc}

\section{Introduction}

The observation of quantum light in phase space is typically associated with homodyne detection \cite{Raymer2009}. 
However, photon counting also provides attractive approaches for studying the distinct quantum character  of the quasi-probability distributions \cite{Cahill1969, Cahill1969a}. 
Using the photon-parity operator, for example, even individual points of the Wigner function can be directly probed despite the Heisenberg's uncertainty relation of the field quadratures \cite{Englert1993, K.Banaszek1996, Wallentowitz1996}.
In the optical regime, this technique has already been utilized  for the characterization of cavity confined optical states \cite{Lutterbach1997, Bertet2002, Lougovski2003, Hofheinz2009} as well as free propagating light fields \cite{K.Banaszek1999, Allevi2009, Bondani2009, Laiho2010}. 
Even though the photon-number parity can be straightforwardly deduced from the measured photon statistics, the detection losses of practical photon counters tend to destroy the real quantum characteristics  featured in the photon-number distribution.
Therefore, sophisticated methods for testing the nonclassical character of the loss-degraded states via their photon-number content have been developed  \cite{Waks2006a, Jezek2011}, and techniques allowing a loss-tolerant reconstruction of  photon statistics are of great importance \cite{Hong1986,  Waks2004, Zambra2005, Achilles2006, Avenhaus2008, Wasilewski2008}.
Still, an accurate determination of the state's properties from the loss-degraded data is challenging, 
especially when the studied state incorporates higher-photon number contributions \cite{K.Banaszek1997,  Achilles2004, Laiho2009}. 


Genuine quantum features can fortunately be recognized even without access to the state's complete phase-space representation. 
One of the pioneering techniques was introduced by Hanbury Brown and Twiss \cite{Hanbury1956}, and in an extended form  \cite{Shchukin2005} their experiment allows us to access the higher order factorial moments  of photon number \cite{Assmann2009, Avenhaus2010, Stevens2010}. 
In general the  $m$-th order normalized factorial moment is determined as 
\begin{equation}
g^{(m)} = \braket{n^{(m)}} / \braket{n}^{m}, 
\label{eq_g}
\end{equation}
where $\braket{n^{(m)}} = \braket{: \hat{n}^{m}:} = \sum_{n} n (n-1)\dots(n-m+1) \ \varrho(n)$  and  $\braket{n} = \braket{n^{(1)}}$ can be  evaluated either as normally ordered ($: :$) moments of the photon number operator $\hat{n}$ or via the photon statistics $\varrho(n)$ with $n$ being the photon number  \cite{L.Mandel1995}. 
Usually, the normalized form of the factorial moments can be extracted loss independently \cite{S.M.Barnett1997, Avenhaus2010}. However, care has to be taken when detecting multimode states since the individual modes may suffer from different amounts of losses \cite{Laiho2011}.
Nonetheless, these moments provide versatile alternatives for investigating the properties of quantum optical states. 


Regarding single photons \cite{Waks2006}, already their second order normalized factorial moment, which ideally takes the value $g^{(2)} = 0$, can be employed as a valuable characterization tool. 
The real measured values are widely used to classify the practical single-photon sources \cite{Eisaman2011}. 
More generally, however, observing $g^{(2)}<1$  can be regarded as a signature of  the nonclassicality  \cite{Korbicz2005} and as an indication of the sub-Poissonian photon-number characteristics \cite{Short1983, Rarity1987, U'Ren2005}. 
Nonetheless, when single photons are displaced  their  $g^{(2)}$ values gradually increase and
finally exceed unity, which signalizes  a super-Poissonian photon-number distribution \cite{Moya-Cessa1995}. 
In this region also a more sophisticated test for the nonclassicality should be found  \cite{Agarwal1992,W.Vogel2008}.  
Further,  once having accessed the photon-number content of displaced single photons  \cite{A.I.Lvovsky2002, Laiho2010} many other intriguing phase-space features  can be directly scrutinized such as the nonclassical oscillations in the photon statistics  \cite{W.Schleich1987, Oliveira1990}. 
Moreover, the factorial moments of \emph{displaced} states indeed provide routes for 
accessing more elaborate moments of the photon  creation and annihilation operators \cite{Shchukin2005}.


Here, we measure up to the fourth order normalized factorial moments  of phase-averaged and displaced single photons directly in a coincidence counting experiment by employing the time-multiplexed detector (TMD)  \cite{Achilles2003, Fitch2003}
that has proven to be a powerful tool  for measuring the higher order moments of  pulsed quantum states of light \cite{Avenhaus2010}.
Even without access to the complete photon statistics we can loss-independently observe quantum features in the prepared states. 
Further, by calibrating the mean photon number of the loss-degraded states we gain information about their photon-number content.
At low detection efficiencies, it is generally a highly nontrivial task to invert the action of losses \cite{Kiss1995}. Further, in order to estimate the reliability of the state reconstruction often numerical methods are applied 
such that the effects of statistical fluctuations, losses, and highest resolved photon number can also be taken into account \cite{Laiho2009, Laiho2010}.
In contrast to the ordinary loss inversion, our technique allows us to study the effect of different experimental limitations separately from each other.
As a consequence, we can directly define the boundaries that an experimental realization sets for the state reconstruction in phase space. 
Further, the artifacts introduced  by the nonideal detection can be recognized in a straightforward manner.

This paper is organized as follows. In Sec.~\ref{sec_gen_function} we review the properties of factorial moments, which are connected to the photon statistics via the moment generating function. 
In Sec. \ref{sec_displaced_single_photon} we survey the properties of displaced single photons and review the effects caused by experimental imperfections. 
In Sec.~\ref{sec_experiment} we utilize the factorial moments to investigate the characteristics of the prepared displaced single photons. 

\section{\label{sec_gen_function}Retrieving state characteristics via factorial moments}

The normalized factorial moments can be directly utilized for different state characterization tasks such as discrimination between sub- ($g^{(2)} < 1$) and super-Poissonian ($g^{(2) }> 1$)  photon-number distributions \cite{Grosse2007, Bartley2012} or classification of different quantum states \cite{Avenhaus2010}.
Additionally, they  provide several alternatives for the direct examination of quantum features \cite{Avenhaus2010, Allevi2011}.
One option for investigating phase-insensitive nonclassical behavior in a single mode 
 is to study whether  it is possible to violate the criterion
\begin{equation}
g^{(m+1)} \ge g^{(m)} \ge 1
\label{eq_classic}
\end{equation}
that classical states obey  \cite{Sudarshan2006},
and we treat the multimode states equivalently.
For this purpose, also the measurement of the normalized factorial moments with orders higher than two becomes relevant when regarding states with super-Poissonian characteristics \cite{Agarwal1992}.

A more detailed investigation of the state's photon-number content  via the factorial moments is possible by employing the moment generating function \cite{S.M.Barnett1997, Barnett1998}
that is described in terms of a real valued variable $0\le \mu \le2$  as 
\begin{equation} 
M(\mu) = \sum_{n}\varrho(n)(1-\mu)^{n}.
\label{eq_MGF}
\end{equation}
At the upper bound, that is when $\mu = 2$,  the moment generating function provides  information on the photon-number parity.
Nevertheless, equally interesting is the lower bound since the derivatives of Eq.~(\ref{eq_MGF}) at $\mu =0$ are directly connected with the factorial moments by 
\begin{equation}
\braket{n^{(m)}} =  \left( -\frac{\textrm{d} }{\textrm{d}\mu}\right)^{m} \left. M(\mu) \right |_{\mu = 0}.
\label{eq_n_m}
\end{equation}

Once having accessed the factorial moments the expression in Eq.~(\ref{eq_MGF}) can be re-written as an expansion 
\begin{equation}
M(\mu) = \sum_{m} \frac{(-1)^{m} \braket{n^{(m)}}}{m!} \mu^{m},
\label{eq_M_taylor}
\end{equation}
and  the photon statistics is again reconstructed from the moment generating function 
in Eq.~(\ref{eq_M_taylor}) by 
\begin{equation}
\varrho(n) = \frac{1}{n!}\left( -\frac{\textrm{d} }{\textrm{d}\mu}\right)^{n} \left. M(\mu) \right |_{\mu = 1} 
                  = \sum_{m\ge n} \frac{(-1)^{m+n}}{n!(m-n)!}  \braket{n^{(m)}},
\label{eq_p_n_rec}
\end{equation}
which obeys the conventional normalization $\sum_{n}\varrho({n})= 1$. Thus, the photon statistics is gained by summing up the factorial moments with proper weight factors.
However, we note that this method for reconstructing the photon statistics can only be successful when the expansion in Eq.~(\ref{eq_M_taylor}) converges near $\mu = 1$.


The loss tolerance  in Eq.~(\ref{eq_p_n_rec}) is achieved after re-writing  $ \braket{n^{(m)}}= g^{(m)}\braket{n}^{m} $ and deducing the mean photon number  $\braket{n} $ from the loss-degraded  measurement via $\braket{n}_{\textrm{lossy}}  = \eta \braket{n}$, in which $\eta$ is  the detection efficiency \cite{S.M.Barnett1997}. 
Thus, apart from recording the different orders of  $g^{(m)}$,  this method further entails a calibration of $\eta$ and measurement of $\braket{n}_{\textrm{lossy}}$.
If  the mean photon number can be calibrated accurately,  the reconstruction of photon statistics via normalized factorial moments becomes especially expedient at low detection efficiencies.
The only constraint lies in measuring enough orders of $g^{(m)}$ with good precision during a finite integration time.
As a consequence, the highest statistically accessible moment, which is known from the experimental data, limits the possibilities to completely reconstruct  the state's photon-number content. 
Natural limits are determined by the physical bounds $0 \le \varrho(n)  \le1$. 
More stringent conditions may be found by investigating the distinctive features in the photon statistics of the studied states.
%

\section{\label{sec_displaced_single_photon}Modeling free-propagating displaced single photons}

Even though displaced single photons have been studied in several experiments \cite{Bertet2002, A.I.Lvovsky2002, Hofheinz2009, Laiho2010}, the generated states are seldom ideal and imperfections in the preparation process have to be taken into account. 
We first investigate the properties of ideal displaced single photons and then regard the effect  of experimental imperfections. 
We consider a displacement with a mode mismatch and take into account higher photon-number contributions of the prepared state.

\begin{figure} \centering
\includegraphics[width =0.63\textwidth]{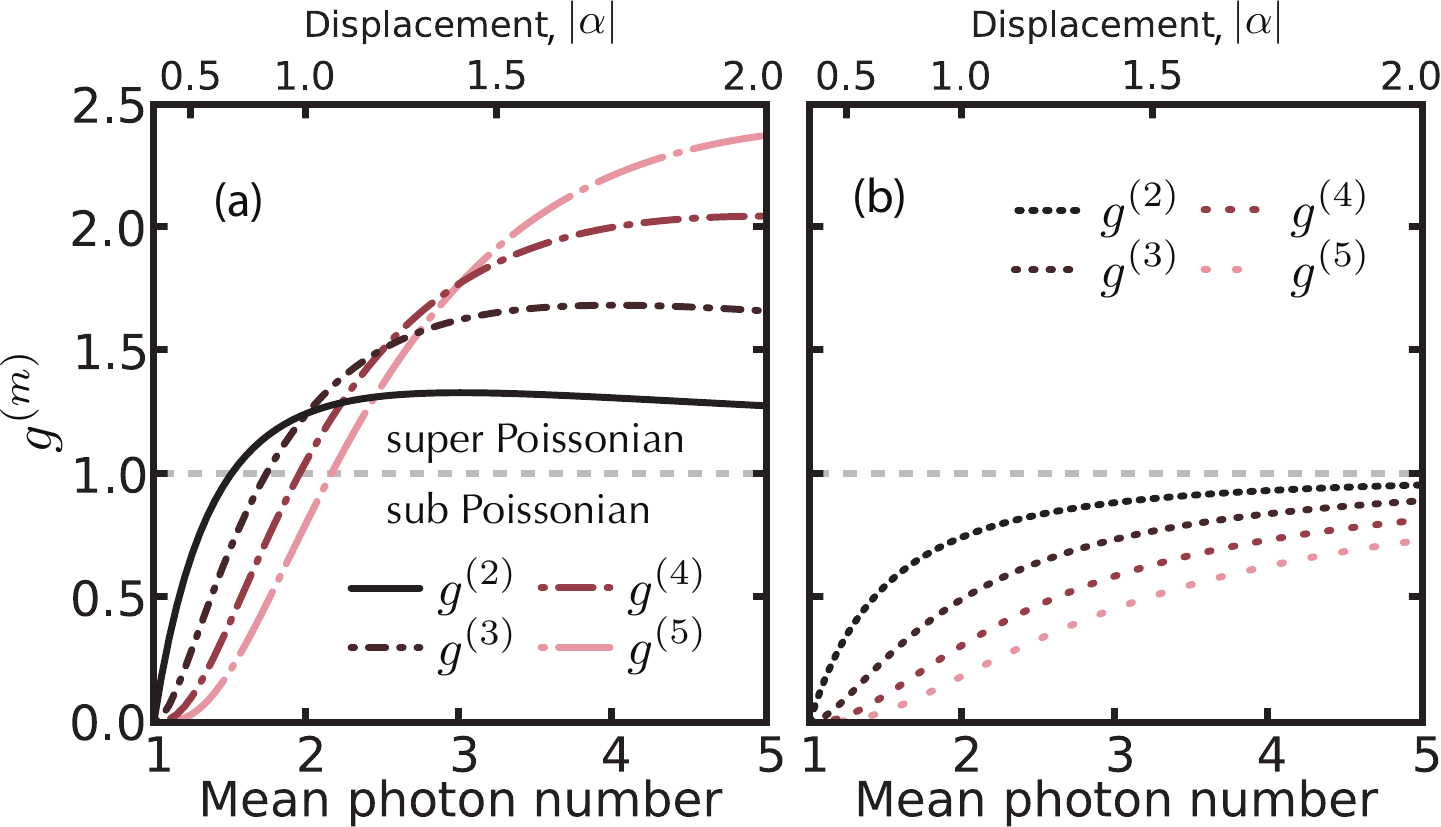}
\caption{Four orders of the normalized factorial moments, shortly $g^{(m)}$ ($m= 2,3,4,5$), with respect to the mean photon number (a) for the ideal displaced single-photon states evaluated according to Eq.~(\ref{eq_g_ideal_fock})  and  (b) in the case of no mode overlap between the single-photon Fock state and the reference state as predicted by Eq.~(\ref{eq_g_two_modes}). Dashed lines separate the sub- and super-Poissonian regions. }
\label{fig_1}
\end{figure}

In the single-mode picture, an ideal displaced single-photon state is described as $\hat{D}(\alpha)\ket{1}$, where  $\hat{D}(\alpha)$ [with the hermitian conjugate $\hat{D}^{\dagger}(\alpha)$] defines the displacement by an amount of $\alpha$ and  $\ket{1} $ is the single-photon Fock state.
We deduce its factorial moments  by evaluating the normally ordered mean values in Eq.~(\ref{eq_g}) with the help of the transformations $\hat{D}^{\dagger}(\alpha)\hat{a}\hat{D}(\alpha) = \hat{a}+\alpha$ and $\hat{D}^{\dagger}(\alpha)\hat{a}^{\dagger}\hat{D}(\alpha) = \hat{a}^{\dagger}+\alpha^{*}$  of the photon annihilation  ($\hat{a}$)  and creation ($\hat{a}^{\dagger}$) operators. 
After a straightforward calculation, the normalized factorial moments of  the ideal displaced single-photon state can be expressed as
\begin{equation}
g^{(m)}_{\textrm{ideal}} = \frac{|\alpha|^{2(m-1)} (m^{2}+ |\alpha|^{2})}{(1+|\alpha|^{2})^{m}}.
\label{eq_g_ideal_fock}
\end{equation}
These moments as depicted in Fig.~\ref{fig_1}(a)  rapidly grow from zero when increasing the mean photon number given by $\braket{n}_{\textrm{ideal}} = 1+|\alpha|^2$.
By following the behavior of the second normalized moment, one directly concludes the gradual transition in the photon-number characteristics of the displaced states \cite{Oliveira1990, Moya-Cessa1995},
and this moment reaches the maximal value of $g^{(2)}_{\textrm{ideal}}  \approx 1.333$ at the displacement  $|\alpha|_{\textrm{max}} = \sqrt{2}$. 
Further, the ideal  displaced single-photon states  always violate the criterion in Eq.~(\ref{eq_classic}). 
However, as seen in Fig.~\ref{fig_1}(a),  the verification becomes more and more challenging  when increasing the displacement since moments with higher and higher orders have to be resolved. 

In practical applications, the displacement can be implemented with an asymmetric beam splitter, at which  the studied state is overlapped with a coherent reference state (see e.g. \cite{Laiho2009} and the references therein). 
The mismatch of the two overlapping modes in temporal, spectral or spatial degrees of freedom can be modeled with a simple overlap factor $\mathcal{M}$. 
In order to evaluate the required mean values in Eq.~(\ref{eq_g}), the photon-number operator can be replaced with 
\begin{eqnarray}
\label{eq_N_eff}
\hat{n} &\rightarrow& \eta  \ \hat{n}_{\textrm{eff}} \\
& = &\eta \left[ \hat{D}^{\dagger}(\sqrt{\mathcal{M}}\alpha)\hat{a}^{\dagger} \hat{a} \hat{D}(\sqrt{\mathcal{M}}\alpha) + (1-\mathcal{M})|\alpha|^{2}\right], \nonumber
\end{eqnarray}
where $\eta$ is the total detection efficiency, $\mathcal{M}$ is the mode overlap,  and $\alpha$  is
the amount of the applied displacement \cite{Laiho2009}. 
The effective photon-number operator $\hat{n}_{\textrm{eff}}$ in Eq.~(\ref{eq_N_eff}) is a sum of a displaced photon-number operator and a background term.

The total detection efficiency cancels out when evaluating the normalized factorial moments, and we can 
write them loss independently in the form
\begin{equation}
g^{(m)}_{\textrm{eff}}  = \frac{\braket{: \hat{n}_{\textrm{eff}}^{m}:}}{\braket{\hat{n}_{\textrm{eff}}}^{m}} = \frac{  \sum_{k}\bigg ( \hspace{-1ex} \begin{array}{c} m \\ k \end{array}  \hspace{-1ex}  \bigg )   \braket{n^{(k)}}_{\textrm{D}}  \braket{n^{(m-k)}}_{\textrm{bg} }}{  (\braket{n}_{\textrm{D}} +  \braket{n}_{\textrm{bg}})^{m}  },
\label{eq_g_two_modes}
\end{equation} 
in which  $ \braket{n^{(m)}}_{\textrm{bg}} = [(1-\mathcal{M})|\alpha|^{2}]^{m}$ describes the properties of the background and 
\begin{eqnarray}
 \braket{n^{(m)}}_{\textrm{D}} &=& \braket{ : [\hat{D}^{\dagger}(\sqrt{\mathcal{M}}\alpha)\hat{a}^{\dagger} \hat{a} \hat{D}(\sqrt{\mathcal{M}}\alpha)]^{m}:} \nonumber \\
 &=& \braket{\hat{D}^{\dagger}(\sqrt{\mathcal{M}}\alpha)\hat{a}^{\dagger m} \hat{a}^{m} \hat{D}(\sqrt{\mathcal{M}}\alpha)}
 \label{eq_O}
 \end{eqnarray}
predicts the behavior of the displaced part. 
If the mode overlap is imperfectly aligned, the results of the measurement change drastically.
As shown in Fig.~\ref{fig_1}(b), in the case of the single-photon Fock state the effective normalized factorial moments cannot take values larger than unity  when $\mathcal{M} = 0$.
Therefore, the super-Poissonian region becomes a loss-independent indicator that the displacement takes place, in other words that $\mathcal{M} \neq 0$.
Nevertheless, our model in Eq.~(\ref{eq_g_two_modes}) provides inaccurate results if only single-photon Fock states are regarded. 
This can be as corrected by considering the higher photon-number contributions of the real single-photon source.
We assume that the single photon is prepared into a photon-number mixed state  \cite{Christ2012},
and we take the higher photon-number contributions of it into account in Eq.~(\ref{eq_O}) when fitting Eq.~(\ref{eq_g_two_modes}) against the  values measured for the displaced single-photon states. 
The overlap factor $\mathcal{M}$ is held as fitting parameter.
Moreover, the amount of displacement applied to the prepared single photon can be straightforwardly extracted in our model from the effective mean photon number. 
Even in the case of imperfect mode overlap this is given by  $ \braket{ \hat{n}_{\textrm{eff}} }= \braket{n}_{\textrm{sp}}+|\alpha|^{2} $, in which $\braket{n}_{\textrm{sp}}$ is the mean photon number of the prepared single photon.

\section{\label{sec_experiment}Experimental investigation of displaced single photons}

\begin{figure} \centering
\includegraphics[width = 0.65\textwidth]{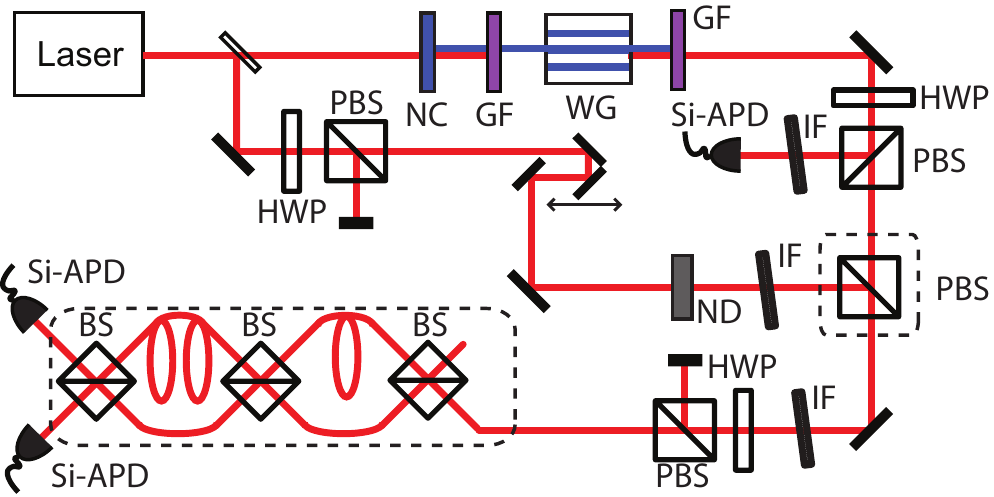}
\caption{Experimental setup in accordance with \cite{Laiho2010}.  
Ti:Sap\-phire laser pulses (wavelength \unit{796}{\nano\metre},  bandwidth \unit{10}{\nano\meter}, repetition rate \unit{2.4}{\mega\hertz})  are frequency doubled in a nonlinear crystal (NC) and coupled to a \unit{1.45}{\milli\meter} long, periodically poled, type-II  $\textrm{KTiOPO}_{4}$ waveguide (WG). 
While color glass filters (GF)  block the residual  beams, in the WG generated  twin beams are separated at a polarizing beam splitter (PBS).  Selected by the orientation of a half-wave plate (HWP), one of them---the herald---is  filtered to a bandwidth of \unit{1}{\nano\meter}  with an interference filter (IF) and sent to a silicon avalanche photodiode (Si-APD). Meanwhile the heralded single photon is coupled to the same spatial and temporal mode with a cross-polarized coherent reference beam attenuated to the single-photon level with a neutral density filter (ND) and filtered to a bandwidth of \unit{1}{\nano\meter}.
The cross-polarized beams are then sent through a spectral filter with a bandwidth of \unit{1}{\nano\meter}. The  displacement  is realized in a phase-averaged manner with a PBS placed after a HWP, whose axis was minimally tilted from the polarization direction of the heralded state. 
Finally, the TMD divides the incoming light pulse in  three subsequent symmetric beam splitters (BS) into eight temporal bins that were detected with two Si-APDs. Components in dashed boxes were implemented  with fiber-integrated optics. }
\label{fig_2}
\end{figure}

In our experiment, shown in Fig.~\ref{fig_2}, we heralded single photons from a pulsed waveguided twin-beam source based on parametric downcoversion. 
For preparing displaced states, the heralded single photons were overlapped at an asymmetric beam splitter with coherent reference states in a phase-averaged manner. 
The amount of displacement was controlled by altering the mean photon number in the reference beam. 
The heralded displaced single photons were then coupled to a TMD for detection. 
Our experimental arrangement is similar to the one used \cite{Laiho2010} except that the TMD is employed for measuring manyfold coincidence counts and not the so-called click statistics. 
We note, however, that the values of the normalized factorial moments could also be deduced via the loss-degraded photon statistics that can be retrieved from the click statistics recorded with TMD by taking into account the intrinsic detector characteristics \cite{Bartley2012}.

As described in \cite{Avenhaus2010}, the higher order correlations in a light beam can be accessed with the TMD detection scheme, and the time-integrated measurement delivers  the desired expectation values \cite{Christ2010}.
Thence, the value of $g^{(m)}$ is extracted from the raw data by dividing the probability to measure a coincidence click between $m$ selected temporal TMD bins by the product of the single click probabilities in these bins---all of them conditioned on the detection of the herald.  
In order to apply the measured normalized moments for the loss-tolerant reconstruction of photon statistics, we further require a calibration of the detection efficiency such that the mean photon number can be determined.
The perfect photon-number correlation between the twin beams empowers us to estimate the detection efficiency according to Klyshko \cite{Klyshko1977}. 
For this purpose we block the reference beam, subtract the amount of accidental counts from the coincidences between the twin beam detection, and compare this number to the amount of single counts in herald. 
The mean photon number of the displaced state merely follows  from dividing the first \emph{un}normalized factorial moment---the probability of measuring a single click conditioned on the detection of the herald---by the estimated detection efficiency.

\begin{figure} \centering
\includegraphics[width = 1.0\textwidth]{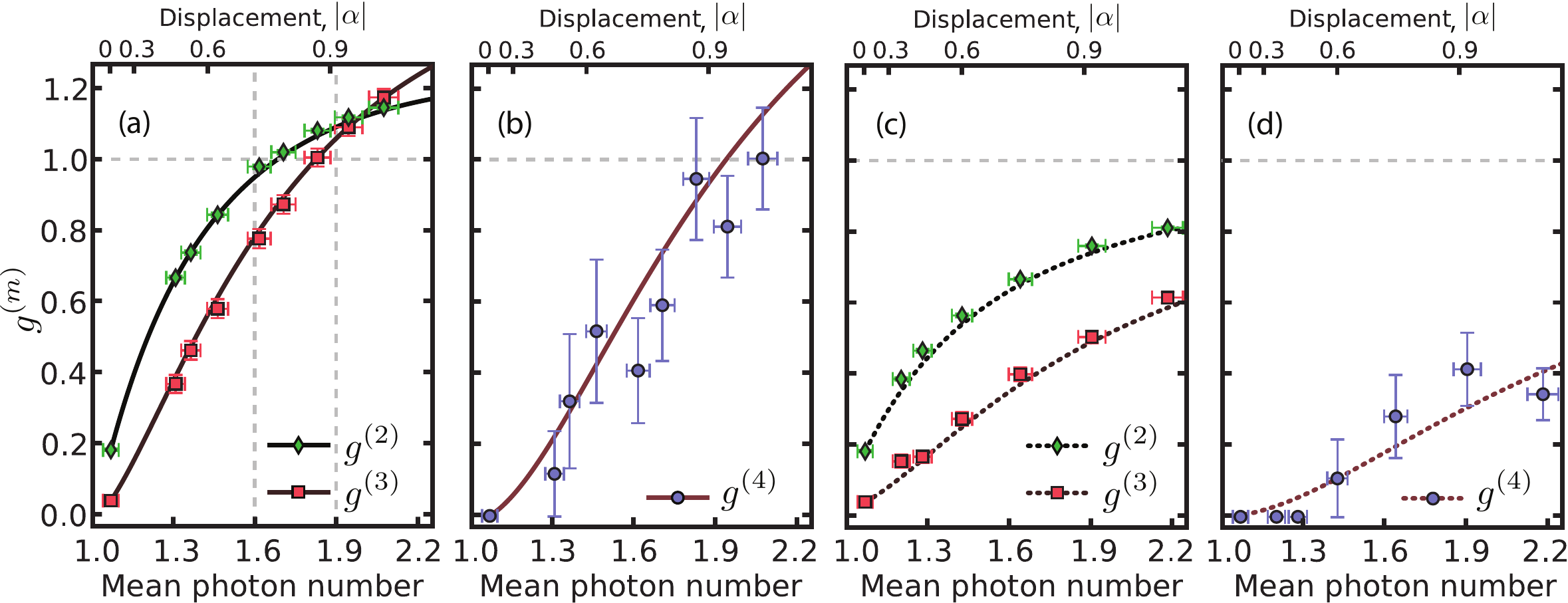}
\caption{The measured values (symbols) of (a) $g^{(2)}$ and $g^{(3)}$ as well as (b) $g^{(4)}$ in dependence of the calibrated mean photon number when the mode overlap was optimized  to $\mathcal{M} = 0.71(2)$. (c) The same as panel (a); and (d) the same as panel (b) for the case of deliberately mismatched mode overlap. 
Solid lines are fitted, whereas dotted lines predict the results for $\mathcal{M} = 0$.   Vertical errorbars are given by statistical fluctuations only, whereas the horizontal ones are dominated by the accuracy of the efficiency estimation. Dashed lines are a guide to the eyes.}
\label{fig_3}
\end{figure}

We first study the values of the moments $g^{(2)}$ to $g^{(4)}$ in order to categorize the photon-number properties of the prepared states.
For the prepared heralded single-photon state we extracted the values  $g^{(2)} = 0.184(4)$ and  $g^{(3)} = 0.04(2)$, the accuracy of which is limited by the statistical fluctuations only.
In contrast to a genuine single-photon state, one clearly recognizes an additional two-photon contribution in the heralded state.
This is the trade-off from a rather high pump power---on average \unit{12}{\micro\watt},  which nevertheless yielded a heralding rate of  \unit{11.8}{\kilo\hertz}. 
When the prepared single photon is now displaced, the values of  $g^{(2)}$  to $g^{(4)}$ gradually increase.
This behavior is shown in Figs~\ref{fig_3}(a-b) with respect to the calibrated mean photon number.
Maximally we observed the values  $g^{(2)}_{\textrm{max}} = 1.148(6)$, $g^{(3)}_{\textrm{max}} = 1.18(3)$ and $g^{(4)}_{\textrm{max}} = 1.00(15)$.
A comparison of these results with the ones gained for the case of vanishing mode overlap [Figs.~\ref{fig_3}(c-d)] reveals, as expected, the rapid appearance of the higher order normalized factorial moments, when the prepared single photon is displaced. 
In contrast to our earlier studies \cite{Laiho2010}, we now fit our model in Eq.~(\ref{eq_g_two_modes}) against the properties of the higher photon-number contributions. 
Our model assumes that all the photon-number components of the prepared single photon possess the same overlap factor and a reasonable match for the data in Figs~\ref{fig_3}(a-b) is found. 
If the overlap factor was drastically lower for the higher photon-number contributions---in our case the two-photon contribution---we would expect to encounter difficulties when fitting our model against the values  measured for the third or higher order normalized factorial moments.

Even without information on the complete photon statistics, we conclude from our results in Fig.~\ref{fig_3}(a)  that the expected transition between the sub- and super-Poissonian photon statistics takes place. 
Further, our  displaced states show quantum features at small values of displacement.
As seen in Fig.~\ref{fig_3}(a), the inequality $g^{(2)} \ngeq1$ guarantees the nonclassicality of the prepared states below the mean photon number of approximately 1.6. 
Our results  in Fig.~\ref{fig_3}(a) clearly also violate the classicality via  $g^{(3)} \ngeq g^{(2)} $. 
However, this criterion becomes inadequate close to the mean photon number of 1.9, beyond  which we are unable to detect quantum features in the prepared displaced states. 
In order to do so, a more accurate measurement of the fourth normalized moment $g^{(4)}$ than the one shown in Fig.~\ref{fig_3}(b) is required.

\begin{figure} \centering
\includegraphics[width = 0.63\textwidth]{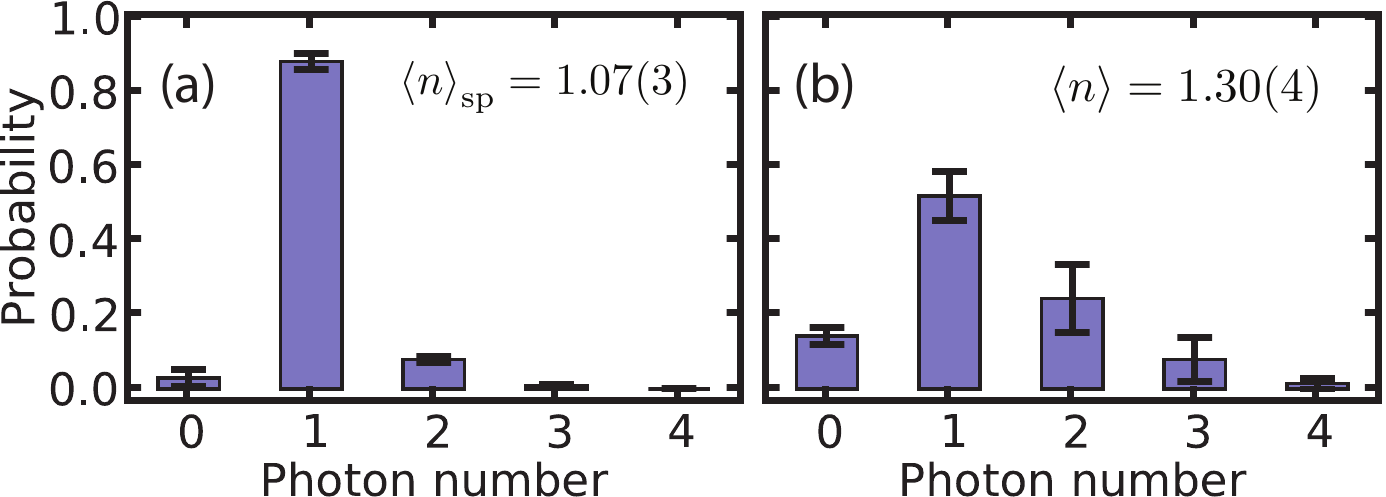}
\caption{The reconstructed photon statistics of  the (a) heralded and (b) minimally displaced single-photon states.} 
\label{fig_4}
\end{figure}

Next, we examine, how much information on the displaced state's photon-number content can be deduced via the measured moments. 
We estimate the photon-number contributions of the heralded and displaced states 
by plugging the measured values in Figs~\ref{fig_3}(a-b) into Eq.~(\ref{eq_p_n_rec}) together with the calibrated mean photon numbers.
For the heralded single photon, the photon statistics of which is shown Fig.~\ref{fig_4}(a), we extracted the loss-calibrated mean photon number of $\braket{n}_{\textrm{sp}} = 1.07(3)$ that was estimated in a measurement with little less than one percent detection efficiency per TMD bin.
Further, we note that the accuracy at which the mean photon number is deduced depends not only on the statistical fluctuations but also on the approximative estimation of the de\-tec\-tion efficiency.
As shown in Fig.~\ref{fig_4}(b), we can recover the behavior expected for  the displaced single photon---increasing vacuum contribution and decreasing one-photon contribution---at the mean photon number of $1.30(4)$. 

\begin{figure} \centering
\includegraphics[width = 0.63\textwidth]{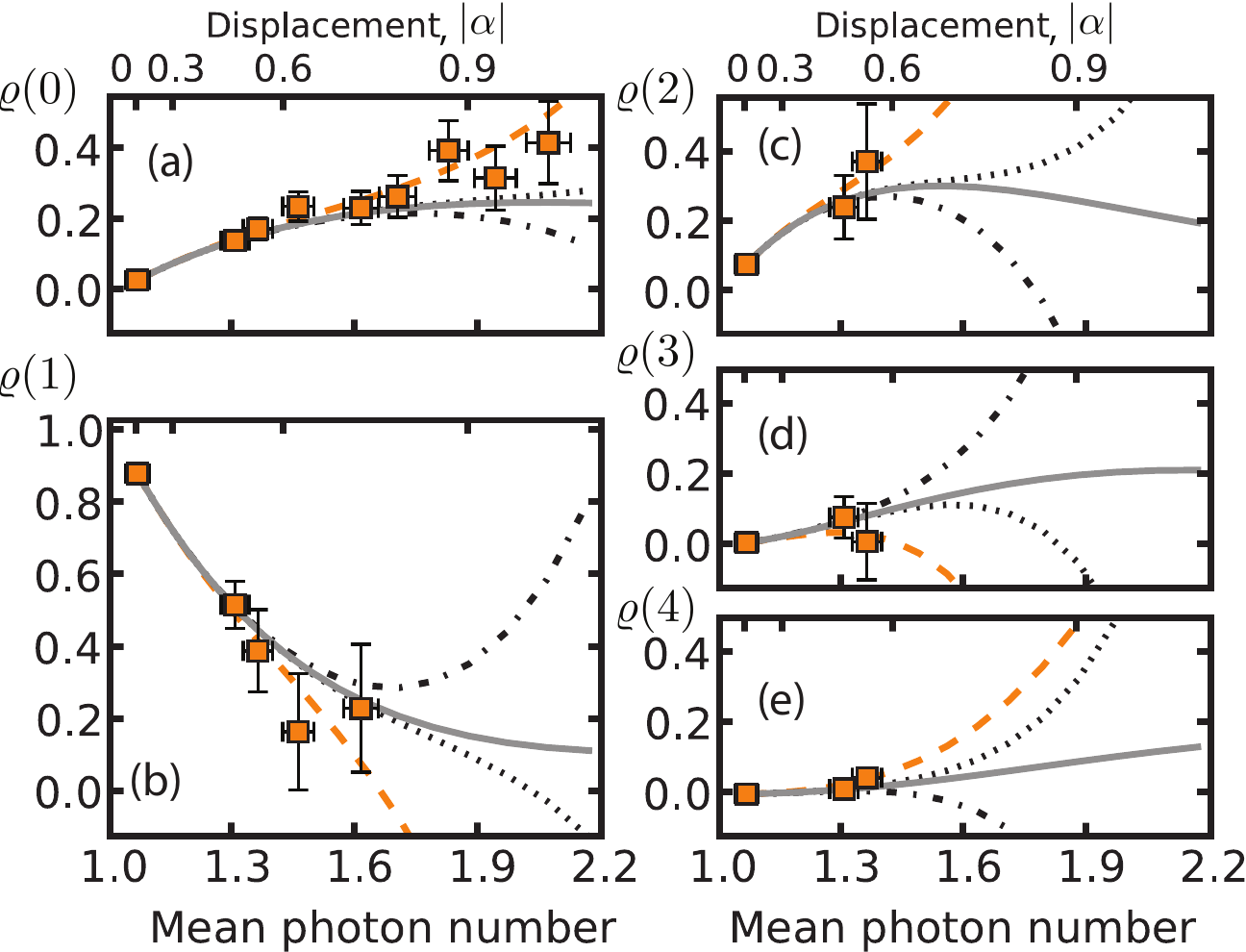}
\caption{The reconstructed photon number contributions $\varrho(0)$--$\varrho(4)$ (a--e) for the displaced states prepared in Figs~\ref{fig_3}(a-b). Symbols originate from measured values. The (orange) dashed lines present the truncation of moments up to $g^{(4)}$ and is gained from the fits in Figs~\ref{fig_3}(a-b), whereas the (black) dash-dotted and dotted lines include the prediction of moments up to $g^{(5)}$ and $g^{(6)}$, respectively. The (gray) solid line represents the theoretically expected photon statistics.} \label{fig_5}
\end{figure}

However, the limited resolution restricts  the acceptable phase-space displacement and sets a boundary to a region, in which the photon statistics of the prepared displaced state can reliably be reconstructed.
Being able to statistically resolve moments only up to the fourth order, in other words the experiment delivers $g^{(n > 4)} = 0$,  the two highest accessible photon-number contributions are deduced  according to Eq.~(\ref{eq_p_n_rec}) as $\varrho(4) = \braket{n}^{4}/4! \ g^{(4)}$  and  $\varrho(3) = \braket{n}^{3} / (3)! \  [g^{(3)}- g^{(4)}\braket{n}]$. 
The non-negativity of the photon-number components provides us with a condition  $ \braket{n} \le g^{(3)}/g^{(4)} $ for the  accessible reconstruction range.
By employing the fits in Figs~\ref{fig_3}(a-b) we get the bound $\braket{n} \lesssim1.4$.
Looking closer at the individual photon-number contributions in Fig.~\ref{fig_5}, a second, stricter condition is obtained by studying the boundary, close to which the reconstructed photon-number contributions start to  significantly deviate from the  expected behavior. 
As the highest resolvable photon-number component, in our case $\varrho(4)$ [Fig.~\ref{fig_5}(e)--dashed line], gradually increases with respect to mean photon number, it  eventually surpasses the contribution $\varrho(3)$ [Fig.~\ref{fig_5}(d)--dashed line]. This is an artifact not expected in the photon statistics of the displaced single photon at the studied region. 
Therefore, we use the limitation $ \varrho{(3)}\ge\varrho{(4)} $ as the boundary of the reliable reconstruction range for the displaced prepared states and gain the ultimatum $\braket{n} \lesssim1.3$.
Furthermore, we note that the vacuum component [Fig.~\ref{fig_5}(a)--symbols] of the  prepared displaced states could be fairly reliably extracted from our measurement  even beyond the determined range, whereas the higher photon-number contributions  [Figs~\ref{fig_5}(b-e)--symbols] are not as resilient.

In order to accurately  reconstruct the complete photon statistics outside the determined region,  one is required to resolve moments with orders higher than four. 
Thus, we predict the values of the moments $g^{(5)}$ and $g^{(6)}$ in order to interpret their influence on the photon statistics. 
As depicted in Fig.~\ref{fig_5} the accessible reconstruction range gradually increases when moments with higher orders are resolved.
Clearly, the method used provides us with insight into the reliability of the state characterization at low detection efficiencies.
In our example it allows us to estimate the bounds of the reconstruction range in phase space and to recognize the artifacts caused by the nonideal detection. 
Systematical deviations in the reconstructed photon statistics become apparent if enough orders of the normalized factorial moments cannot  be measured due to the limited detection time.
Further, the  statistical fluctuations in the measured moments and the precision at which the mean photon number is extracted affect the accuracy of the reconstructed photon statistics.
Fortunately, their effect can be investigated separately from the systematical deviations and in our case the errorbars are  dominated by the fluctuations in the resolved moments rather than by the inaccuracy in the mean photon number.
In summary, our results can give direct specifications for the experimental parameters when studying the fine structure in  the photon statistics of displaced single photons at low detection efficiencies.

\section{Conclusions}
We measured up to the fourth order normalized factorial moments of displaced single photons in a loss-independent manner. 
By studying the second normalized moment, we confirmed as expected that the sub-Poissonian photon-number distribution of a single photon gradually moves toward the super-Poissonian photon statistics when the state is displaced.
The prepared displaced states further violated the classicality even in the super-Poissonian regime.
Moreover, the measured moments provide means for the loss-tolerant reconstruction of photon statistics after determining the mean photon number of the studied state. 
However, it is essential to accurately measure enough orders of the factorial moments in order to reliably reconstruct the investigated properties.
Our results show the versatility of the factorial moments for the state characterization,
and we believe they prove to be useful in examining genuine quantum features at low detection efficiencies.

\section*{Acknowledgments}
We thank G. Harder for fruitful discussions.
This work was supported by the EC under the grant agreements Corner (FP7-ICT-213681) and Q-Essence (248095).

\section*{References}
\vspace{3mm}
\bibliographystyle{unsrt}
\bibliography{references}

\end{document}